\documentclass[english]{article}
\usepackage[T1]{fontenc}
\usepackage[latin9]{inputenc}
\usepackage[a4paper]{geometry}
\usepackage{url}
\usepackage{amsmath}
\usepackage{amssymb}
\usepackage{graphicx}
\usepackage{setspace}
\usepackage{cite}
\usepackage[unicode=true]{hyperref}

\makeatletter
\DeclareMathOperator{\rmd}{d}
\DeclareMathOperator{\real}{Re}
\newcommand{\iu}{\mathrm{i}}
\makeatother

\begin{document}
\title{Immediate recapture in the trapping-detrapping process of a single
charge carrier}
\author{Aleksejus Kononovicius\thanks{email: \protect\href{mailto:aleksejus.kononovicius@tfai.vu.lt}{aleksejus.kononovicius@tfai.vu.lt};
website: \protect\url{https://kononovicius.lt}}, Bronislovas Kaulakys}
\date{Institute of Theoretical Physics and Astronomy, Vilnius University}
\maketitle

\begin{abstract}
Previously we have shown that pure 1/f noise arises from the trapping-detrapping
process when traps are heterogeneous. Namely, the trapping-detrapping
process relies on the assumption that detrapping rates of individual
trapping centers in the condensed matter are random and uniformly
distributed. Another assumption underlying the trapping-detrapping
process was that both trapping and detrapping times need to have non--zero
duration. Here we violate the latter assumption by introducing immediate
recapture of the charge carrier. We show that 1/f noise will still
be observed, though the range of frequencies over which it will be
observed shifts to the lower frequency range as the immediate recapture
probability increases.
\end{abstract}

\section{Introduction}

The white noise and the Brownian noise are two most well understood
examples of noise and fluctuations in the various materials and devices
\cite{Kogan1996CUP}. White noise most typically arises from thermal
fluctuations, or the discrete nature of detected particles (i.e.,
shot noise). It is characterized by an absence of temporal correlations,
and a flat power spectral density (abbr. PSD). The Brownian noise
is a temporal integral of the white noise, and consequently exhibits
no correlations between the increments of the signal. The Brownian
noise is short--term correlated, and exhibits PSD of the $S\left(f\right)\sim1/f^{2}$
form. Yet in various materials and devices, especially in the low
frequency range, PSD of $S\left(f\right)\sim1/f^{\beta}$ (with $0.5<\beta<1.5$)
form is often observed. The nature of this noise, in the literature
often referred to as $1/f$ noise, flicker noise, or pink noise, remains
an open question \cite{Voss1976PRL,Dutta1981RMP,Kobayashi1982BioMed,Li2005PRE,Levitin2012PNAS,Fox2021Nature,Wirth2021IEEE}.

Here we extend a model of $1/f$ noise based on the trapping--detrapping
process in the condensed matter \cite{Kononovicius2023PRE,Kononovicius2023upoiss}.
Unlike in numerous previous works (e.g., \cite{Bernamont1937ProcPhysSoc,VanDerZiel1979AEEP,Milotti2002,Palenskis2015SRep})
$1/f$ noise in this particular process arises not from the superposition
of independent relaxation processes, but from a drift of a single
charge carrier. The quick overview of this model is given in Section~\ref{sec:Trapping=002013detrapping-process}.
In Section~\ref{sec:Immediate-recapture} we introduce immediate
recapture mechanism, which violates a core assumption underlying the
original trapping--detrapping process. The recapture mechanism allows
to have zero trapping times, and thus effectively implements ``touching''
gaps. In Section~\ref{sec:Immediate-ejection} we examine similar
mechanism, which allows to have zero detrapping times, and thus effectively
implements ``touching'' pulses. Analytically we show that these mechanisms
do not prevent observation of $1/f$ noise, they just change the range
of frequencies over which pure $1/f$ noise is observed. Outside this
range PSD either saturates (for lowest frequencies) or decays as the
Brownian noise. We supplement our analytical derivations by conducting
numerical simulations. The results are summarized and conclusions
are provided in Section~\ref{sec:Conclusions}.

\section{Trapping--detrapping process\label{sec:Trapping=002013detrapping-process}}

\begin{figure}
\noindent \begin{centering}
\includegraphics[width=0.4\textwidth]{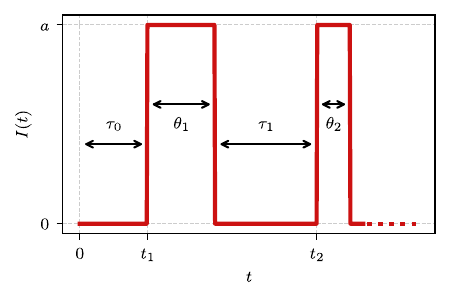}
\par\end{centering}
\caption{Sample signal generated by the trapping--detrapping of a single charge
carrier. Relevant notation: $\tau_{i}$ -- detrapping time (gap duration),
$\theta_{i}$ -- trapping time (pulse duration), $a$ -- current
generated by a free--drifting charge carrier (pulse height).\label{fig:explanation}}
\end{figure}

In contrast to the numerous previous works \cite{Bernamont1937ProcPhysSoc,VanDerZiel1979AEEP,Milotti2002,Palenskis2015SRep},
which consider superposition of independent relaxation processes,
let us consider trapping--detrapping process of a single charge carrier
(e.g., electron) drifting through the condensed matter. In what follows
we assume that the charge carrier can either drift through the conduction
band (thus generating electric current), or be trapped within a trapping
center (thus no current is observed). Under these assumptions, the
electric current (the observed signal) generated by the charge carrier
will be a sequence of non--overlapping rectangular pulses. A sample
of such signal is shown in Fig.~\ref{fig:explanation}.

In Fig.~\ref{fig:explanation}, and further, $\tau_{i}$ stands for
$i$-th detrapping time, $\theta_{i}$ stands for $i$-th trapping
time (when written without the indices $\tau$ and $\theta$ will
stand for the detrapping and trapping times in general). Note that,
detrapping and trapping times are random variates sampled from the
preselected distributions. On the other hand, we assumed that value
of $a$ is predetermined and fixed through the simulation.

The process will generate current $I\left(t\right)$ composed of non--overlapping
rectangular pulses with profiles $A_{k}\left(t\right)$. PSD of such
signal is given by
\begin{equation}
S\left(f\right) =\lim_{T\rightarrow\infty}\left\langle
\frac{2}{T}\left|\int_{0}^{T}I\left(t\right)e^{-2\pi\iu ft}\rmd
t\right|^{2}\right\rangle = \lim_{T\rightarrow\infty}\left\langle
\frac{2}{T}\left|\sum_{k}e^{-2\pi\iu ft_{k}}\mathcal{F}\left\{
    A_{k}\left(t-t_{k}\right)\right\} \right|^{2}\right\rangle .
\end{equation}
In the above $T$ stands for the observation time, $t_{k}$ stands
for time when $k$-the pulse starts, while $\mathcal{F}\left\{ A_{k}\left(t-t_{k}\right)\right\} $
is the Fourier transform of the $k$-th pulse profile. Given the details
of the process considered here the pulses will differ only by their
duration. As the individual trapping and detrapping times are assumed
to be independent, the PSD of $I\left(t\right)$ is determined purely
by pulse height and the trapping and detrapping time distributions
(let $\chi_{\theta}\left(\theta\right)$ and $\chi_{\tau}\left(\tau\right)$
be the respective characteristic functions). Under these considerations
the PSD is given by \cite{Kononovicius2023PRE}
\begin{equation}
S\left(f\right)=\frac{a^{2}\bar{\nu}}{\pi^{2}f^{2}}\real\left[\frac{\left(1-\chi_{\theta}\left(f\right)\right)\left(1-\chi_{\tau}\left(f\right)\right)}{1-\chi_{\theta}\left(f\right)\chi_{\tau}\left(f\right)}\right],\label{eq:rectangular-psd}
\end{equation}
with $\bar{\nu}$ being the mean number of pulses per unit time.

Typically when trapping--detrapping processes are considered \cite{Mitin2002}
it is assumed that both $\tau_{i}$ and $\theta_{i}$ are sampled
from the same distribution. Most often an exponential distribution
is used with rates $\gamma_{\tau}$ and $\gamma_{\theta}$ respectively.
This model would produce a Lorentzian PSD \cite{Mitin2002}.

\begin{figure}
\noindent \begin{centering}
\includegraphics[width=0.4\textwidth]{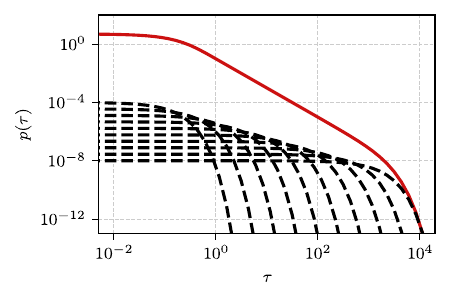}
\par\end{centering}
\caption{Probability density function of the detrapping time distribution under
the assumption that individual detrapping rates are uniformly distributed
in $\left[\gamma_{\text{min}},\gamma_{\text{max}}\right]$ (red curve),
Eq.~\eqref{eq:escape-time-pdf}. The probability density function
was obtained with $\gamma_{\text{min}}=10^{-3}$ and $\gamma_{\text{max}}=10$.
Black dashed curves correspond to the exponential probability density
functions with fixed rates: $\gamma_{\tau}=10^{-3}$, $2.78\times10^{-3}$,
$7.74\times10^{-3}$, $2.15\times10^{-2}$, $5.99\times10^{-2}$,
$1.67\times10^{-1}$, $4.64\times10^{-1}$, $1.29$, $3.59$, and
$10$. Normalization of the exponential probability density functions
was adjusted for the visualization purposes, but it remains proportional
to their respective contributions.\label{fig:distribution}}
\end{figure}

Instead let us assume that the detrapping rates $\gamma_{\tau}$ are
uniformly distributed in $\left[\gamma_{\text{min}},\gamma_{\text{max}}\right]$.
Implying that individual capture centers are heterogeneous, and are
characterized by their own individual detrapping rate. In this case
the distribution of the detrapping times is a continuous mixture of
exponential distributions with different values of the rate parameter
(see Fig.~\ref{fig:distribution}). Then the probability density
function (abbr. PDF) of detrapping time $\tau$ is given by:
\begin{equation}
p\left(\tau\right) =\frac{1}{\gamma_{\text{max}}-\gamma_{\text{min}}}\int_{\gamma_{\text{min}}}^{\gamma_{\text{max}}}\gamma_{\tau}\exp\left(-\gamma_{\tau}\tau\right)\rmd\gamma_{\tau}
=\frac{\left(1+\gamma_{\text{min}}\tau\right)\exp\left(-\gamma_{\text{min}}\tau\right)-\left(1+\gamma_{\text{max}}\tau\right)\exp\left(-\gamma_{\text{max}}\tau\right)}{\left(\gamma_{\text{max}}-\gamma_{\text{min}}\right)\tau^{2}}.\label{eq:escape-time-pdf}
\end{equation}
For very short $\tau$ the PDF saturates, as does the exponential
distribution for very short times. For extremely long $\tau$ the
PDF also decays as an exponential function. For the intermediate values,
$\frac{1}{\gamma_{\text{max}}}\ll\tau\ll\frac{1}{\gamma_{\text{min}}}$,
power--law asymptotic $\tau^{-2}$ behavior is observed. Having $p\left(\tau\right)\sim\tau^{-2}$
is known to be one of the ingredients needed to obtain $1/f$ noise
\cite{Margolin2006JStatPhys,Lukovic2008JChemPhys,Niemann2013PRL,Kononovicius2023PRE}.
In our particular case, it can be shown that for $\gamma_{\text{min}}\ll2\pi f\ll\gamma_{\text{max}}$
pure $1/f$ noise is observed \cite{Kononovicius2023upoiss}
\begin{equation}
S\left(f\right)\approx\frac{a^{2}\bar{\nu}}{\gamma_{\text{max}}f}.\label{eq:main-psd}
\end{equation}

\section{Immediate recapture mechanism\label{sec:Immediate-recapture}}

Now let us assume that immediately after the charge carrier escapes,
it may become immediately trapped with finite probability $r$. This
implies that the trap is escaped after certain random number of attempts.
Let $k$ be the number of attempts. It is trivial to show that $k$
follows geometric distribution, whose probability mass function is
given by
\begin{equation}
p_{\text{geom}}\left(k\right)=r^{k-1}\left(1-r\right).
\end{equation}
As charge carrier has made $k$ attempts to escape the trap, it has
spent $\tau_{i}=\sum_{j=1}^{k}s_{j}$ time being trapped. Here $s_{j}$
would correspond to the detrapping time discussed in the previous
section, but with a caveat that the next trapping time was zero. Instead
of dealing with zero duration trapping times, let us simply consider
that detrapping times are sampled from a different distribution, which
takes into account the immediate recapture.

As immediate recapture is being made by the same trapping center,
we assume that $s_{j}$ have same characteristic rate $\gamma_{\tau_{i}}$.
Under these assumptions, $\tau_{i}$ for a fixed $k$ follows Erlang
distribution. The characteristic function of Erlang distribution is
given by
\begin{equation}
\chi_{\text{Erlang}}\left(f\right)=\left(\frac{\gamma_{\tau}}{\gamma_{\tau}-2\pi\iu f}\right)^{k}.
\end{equation}
Consequently the characteristic function of detrapping time distribution
will be given by
\begin{equation}
\chi_{\tau}\left(f\right) =\sum_{k=1}^{\infty}\left[p_{\text{Geom}}\left(k;r\right)\chi_{\text{Erlang}}\left(f\right)\right]
 =\frac{\gamma_{\tau}\left(1-r\right)}{\gamma_{\tau}\left(1-r\right)-2\pi\iu f}.\label{eq:char-fun-immediate-recapture-case}
\end{equation}
As can be seen from the above the immediate recapture mechanism simply
rescales the rates by a factor of $1-r$. This observation allows
us to reuse earlier results reported in \cite{Kononovicius2023PRE,Kononovicius2023upoiss}
simply by rescaling the rates, then:

\begin{equation}
S\left(f\right)\approx\frac{a^{2}\bar{\nu}}{\gamma_{\text{max}}\left(1-r\right)f}.\label{eq:recapture-psd}
\end{equation}
The approximation will hold as long as the characteristic trapping
time is comparatively long $\gamma_{\theta}\ll\gamma_{\text{max}}\left(1-r\right)$,
though the range of frequencies for which the approximation holds
will shift towards the lower frequencies, and will apply to $\gamma_{\text{min}}\left(1-r\right)\ll2\pi f\ll\text{\ensuremath{\gamma_{\text{max}}}\ensuremath{\left(1-r\right)}}$
range.

\begin{figure}
\noindent \begin{centering}
\includegraphics[width=0.4\textwidth]{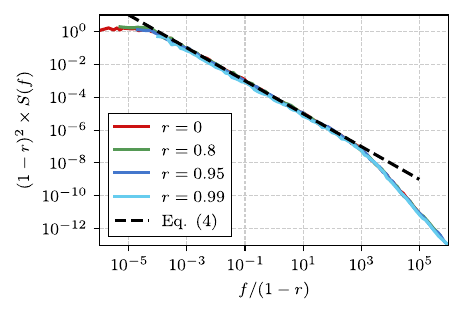}
\par\end{centering}
\caption{Rescaled simulated PSD curves of the trapping--detrapping process
with immediate recapture mechanism. Common simulation parameters:
$T=10^{6}$, $\gamma_{\text{min}}=10^{-4}$, $\gamma_{\text{max}}=10^{4}$,
$a=1$, $\gamma_{\theta}=1$.\label{fig:recapture-psd}}
\end{figure}

The analytical intuition above can be further supported by the numerical
simulation shown in Fig.~\ref{fig:recapture-psd}. In Fig.~\ref{fig:recapture-psd}
power spectral densities obtained for different immediate recapture
probabilities are shown rescaled in a manner that they would fall
on the same curve, which corresponds to the original (no immediate
recapture) case. Based on Eq.~\eqref{eq:char-fun-immediate-recapture-case}
we infer that the frequencies need to be divided by a factor of $1-r$,
while the power spectral density itself needs to be multiplied by
a factor of $\left(1-r\right)^{2}$ due to the mathematical form of
Eq.~\eqref{eq:recapture-psd}:
\begin{equation}
\left(1-r\right)^{2}\times S\left(f\right)\approx\frac{a^{2}\bar{\nu}}{\gamma_{\text{max}}\frac{f}{\left(1-r\right)}}.
\end{equation}

\section{Immediate ejection mechanism\label{sec:Immediate-ejection}}

The general expression for the PSD of a signal with non--overlapping
rectangular pulses, Eq.~\eqref{eq:rectangular-psd}, is symmetric
in respect to the trapping and detrapping time distributions. Namely,
if $\chi_{\tau}\left(f\right)$ and $\chi_{\theta}\left(f\right)$
would be swapped in Eq.~\eqref{eq:rectangular-psd}, then the expression
remain the same. The assymetry is introduced into the model when different
assumptions are being made about the distributions of the trapping
and detrapping times. Though if we swap the assumptions (i.e., sample
detrapping times from exponential distribution, and sample trapping
times from Eq.~\eqref{eq:escape-time-pdf}), the physical interpretation
of the model would change, but the expression for the PSD would remain
the same. In this section, instead of swapping the assumptions and
studying the implications of immediate recapture mechanism again,
let us consider it as an immediate ejection mechanism within the framework
of the original model \cite{Kononovicius2023PRE,Kononovicius2023upoiss}.

Thus the model with immediate ejection, would have $\tau$ being sampled
from a distribution whose PDF is given by Eq.~\eqref{eq:escape-time-pdf},
and $\theta$ being sampled from an exponential distribution with
rate $\gamma_{\theta}$. Implementation of immediate ejection mechanism
would mirror immediate recapture: as the charge carrier is captured
by the capture center, it could be immediately released (ejected)
with probability $r^{\prime}$. Following the same logic as for the
immediate recapture mechanism, we obtain:
\begin{align}
\chi_{\theta}\left(f\right) & =\frac{\gamma_{\theta}\left(1-r^{\prime}\right)}{\gamma_{\theta}\left(1-r^{\prime}\right)-2\pi\iu f}.
\end{align}

The approximation of PSD for the original model, Eq.~\eqref{eq:main-psd},
does not explicitly depend on $\gamma_{\theta}$. It is hidden behind
the mean number of pulses per unit time:
\begin{equation}
\bar{\nu}=\frac{1}{\left\langle \theta\right\rangle +\left\langle \tau\right\rangle }.
\end{equation}
If the trapping times are long in comparison to the detrapping times,
i.e., $\left\langle \theta\right\rangle \gg\left\langle \tau\right\rangle $,
then $\bar{\nu}\approx\left(1-r^{\prime}\right)\gamma_{\theta}$,
while for the original model we would have $\bar{\nu}\approx\gamma_{\theta}$.
The long trapping times assumption is not as restrictive as it may
seem, because it is already known that pure $1/f$ noise can be observed
only with long trapping times \cite{Kononovicius2023PRE,Kononovicius2023upoiss}.
Therefore for the model with immediate ejection, Eq.~\eqref{eq:main-psd}
should hold assuming that $\bar{\nu}$ is calculated appropriately.

In order to make the PSD curves corresponding to the cases with the
different immediate ejection probabilities to fall on the same curve
we need to scale the obtained PSDs by dividing every PSD curve by
a factor of $1-r^{\prime}$. As expected, the simulated PSD curves
fall on the same black dashed curve, which corresponds to Eq.~\eqref{eq:main-psd}.
Though the numerical simulations contradict analytical intuition by
showing that the range of frequencies over which Eq.~\eqref{eq:main-psd}
applies shrinks.

\begin{figure}
\noindent \begin{centering}
\includegraphics[width=0.4\textwidth]{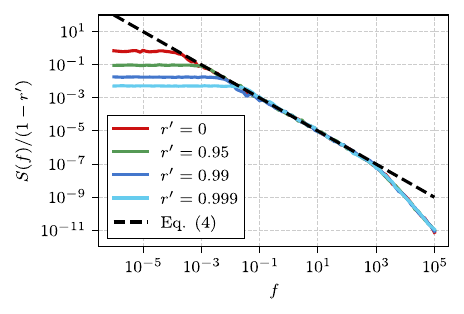}
\par\end{centering}
\caption{Rescaled simulated PSD curves of the trapping--detrapping process
with immediate ejection mechanism. Common simulation parameters: $T=10^{6}$,
$\gamma_{\text{min}}=10^{-4}$, $\gamma_{\text{max}}=10^{4}$, $a=1$,
$\gamma_{\theta}=1$.\label{fig:ejection-psd}}
\end{figure}

\begin{figure}
\noindent \begin{centering}
\includegraphics[width=0.4\textwidth]{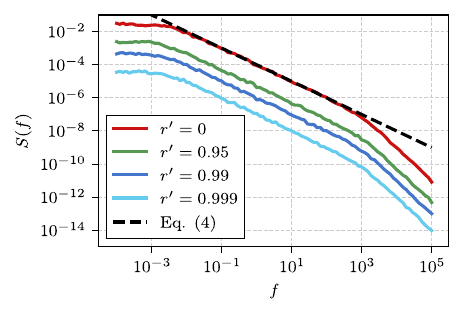}
\par\end{centering}
\caption{Simulated conditional PSD curves of the trapping--detrapping process
with immediate ejection mechanism. Common simulation parameters: $T_{\text{min}}=10^{6}$,
$N_{\text{min}}=10^{6}$, $\gamma_{\text{min}}=10^{-2}$, $\gamma_{\text{max}}=10^{4}$,
$a=1$, $\gamma_{\theta}=1$.\label{fig:ejection-conditional-psd}}
\end{figure}

The shrinkage is caused by the fixed duration of the experiments.
To verify this intuition, let us run numerical simulations, which
are required to contain at least $N_{\text{min}}$ number of pulses,
and have longer duration than $T_{\text{min}}$. Namely, the simulation
is continued until both minimums are exceeded. Our simulations are
similar to the conditional PSD measurements carried out in \cite{Leibovich2017PRE},
but instead of scrapping the experiments we let them continue until
both minimum conditions are met. As shown in Fig.~\ref{fig:ejection-conditional-psd}
the overall shape of PSD is the same in all cases, only the intensity
differs, as no rescaling is applied in this figure.

\section{Conclusions\label{sec:Conclusions}}

We have examined the influence of the immediate recapture mechanism,
and its mirror mechanism -- the immediate ejection, on the spectral
properties of the trapping--detrapping process exhibiting $1/f$
noise. We have shown that making the pulses, or the gaps, disappear
doesn't cause $1/f$ noise to become perverted as taking the point
process limit of the process does \cite{Kononovicius2023PRE,Kononovicius2023upoiss}.
After appropriate rescaling of the signal intensity and the time scale,
both mechanisms still produce PSDs well approximated by Eq.~\eqref{eq:main-psd}.

However, for the particular case of the immediate ejection mechanism,
we have observed spurious shrinkage of the range of frequencies over
which $1/f$ noise is observed. We have shown that the shrinkage is
caused by the decreasing number of pulses being observed over the
finite--duration experiments. This effect disappears if the conditional
PSDs are measured. The simulated experiment is then required to have
a certain minimum duration and to record a certain minimum number
of pulses.

The ideas presented here can be taken further by allowing the individual
pulses to not only ``touch'', but also overlap. However, this will
require new physical and mathematical formulation of the trapping--detrapping
process. In particular, $\tau$ would have to be allowed to become
negative, or its meaning would have to be redefined. Another possible
future research direction would be examining the pulses with arbitrary
profiles.

\begin{singlespace}
\section*{Author contributions}

{\bfseries AK}: Conceptualization, Methodology, Software, Writing,
Visualization. {\bfseries BK}: Conceptualization, Supervision.


\end{singlespace}

\end{document}